# LOW ENERGY ARCHITECTURE IN THE TROPICS : FROM DESIGN TO BUILDING CONTRUCTION

Harry BOYER, Frédéric MIRANVILLE
Réunion Island University
PIMENT Laboratory
117 rue du général Ailleret
97430 Le Tampon
Réunion Island – France (overseas)
harry.boyer@univ-reunion.fr

François PAYET
Ingénieur architecte
22, avenue du 22 juillet 1789, appt. 4
97420 Le Port
Réunion Island – France (overseas)
payet.francois-archi@wanadoo.fr

**ABSTRACT**

This paper presents a realization of passive construction in Reunion, French Department in the Indian Ocean (southern hemisphere) submitted to a wet tropical climate. Aspects of passive construction were integrated at the design stage and this house does not present additional costs towards classical ones. This project was awarded a prize for architecture climate [1].

**KEY WORDS**
Low energy architecture, passive design, energy conscious design.

## 1. Introduction

Passive building design and techniques are well covered in various books [2-5]. Buildings that are really designed in link with climate are not so common in southern hemisphere and this paper presents one real case building. This building is settlement is in the southwest area in Reunion Island, French overseas department. This part of the island enjoys sunshine and trade winds from east to west, particularly in winter. On south side of this project, an enjoyable view was offered on the coast. The climatic characteristics of the site are highly shined (2500 hours of sunshine annually, more than 4.5 kWh / m² daily annual average) and the presence of quite strong trade winds in austral winter. This led mainly to consider a summer design (for much of the year), but also aspects related to the winter design (free gains, ...) have to be taken into account.

## 2. Overview of the project :

Our choice was thus focused on a plot facing east / west, providing a significant linear south side (ocean view), and large areas for solar captation (for the use of solar energy) and convenient orientation relative to ventilation.

The land selected was substantially rectangular, about 700 sqm (approx. 23m * 33m), and from East and West.

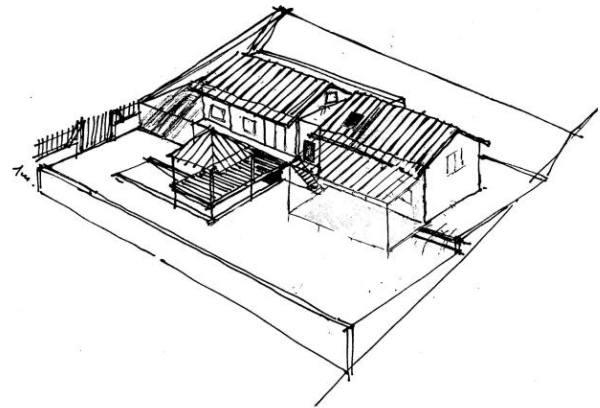

Fig 1 : Architectural drawing of the project

A major constraint was also a significant field gradient (20%), which led us to define several levels on the plot to minimize retaining structures (lower part, southern border), earth moving and visual impact the frame. The program is a house of 170 m² with 4 bedrooms, a living room open to the outside, a garage and an office large enough away from the area of rooms. Our project was that of a construction adapted to the climate and land, with architecture choices rather simple in link with local know how. A significant view being offered on the southern side and this benefit the owner was an important goal. Finally, a program objective was also to try to incorporate as much as possible the construction in the visual environment of the site. The plot is located on the edge of subdivision, the only trees on the south side have been preserved and serve as privacy and sun protection (with regard to surrounding buildings).

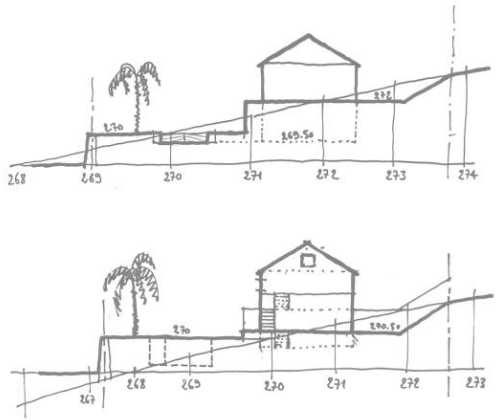

Fig 2 : Drawing Cut view of the project

Depending on the desired technical specifications, rules, costs, orientation, the volume is that of two main blocks with a two slopes roof of very similar appearance (height, shape, roof colors, ...) separated by a space slightly shrinkage in height and the extension of the façade (Fig. 1). The two main blocks are respectively the day (East) and the night one (West). They each have two levels : day part and office for the first), garage and rooms for the second. The day area includes a day room (with kitchen) to low level, extended by a veranda overlooking the garden (Fig. 2). Above a mezzanine area hosts an office and a guest room separated by a bridge.

The connection zone between these two main blocks, is covered with wooden cladding facade, houses the main entrance, stairs, laundry, bathroom and the roof terrace for solar water heaters. The constitution is classical block masonry (concrete block), coatings (single layer coated sand-colored exterior and plaster inside).

Finally ending with an advanced kiosk is disposed at the front of the building and has several functions: opening the bedroom area on a terrace with an unobstructed view of the coast, covering the area of the garage entrance, "window dressing" the view from the entrance while leaving the front of the house a private space. The booth also formed a tag closing the perspective of the access road and an element of comfort in relation to the occupation of the garden. This recalls a vernacular architecture component, the traditional *guetali* allowing to look around the house.

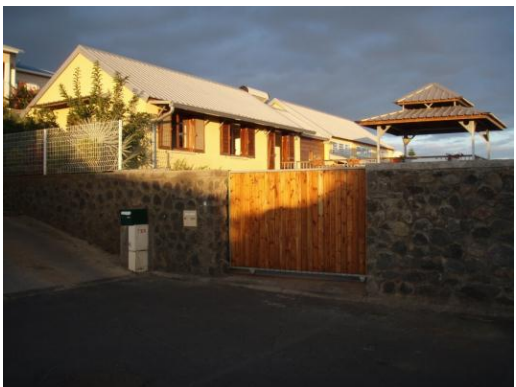

Picture 1 : Global view of the house

## 3. Site integration :

The majority of parcels in the subdivision were oriented N/S and did not allow, according to our criteria, to optimize a design project in relation to aspects of view benefit, ventilation and renewable energy. Our choice was different :

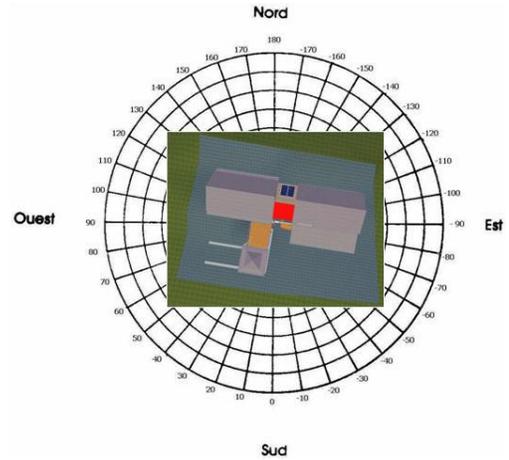

Fig 3 : Implantation of the project

This choice led to minimize East and West surfaces that are difficult to protect from direct solar radiation and also to favorable orientation relative to the ventilation from east to west during the day (trade winds) and North to South during night.

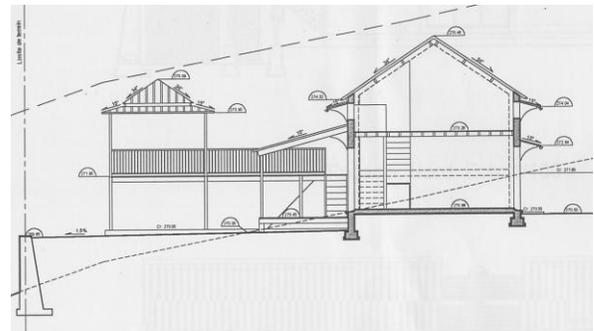

Fig. 4 : Final project cut

## 4. Solar protection :

### 4.1 Roof and walls :

The solar protection is ensured firstly by roof insulation (including large inclined parts facing north and strongly exposed). This isolation makes use of radiant barriers [13]. Thermal simulations were conducted using a validated software specially in link with these aspects [6-12]. The installation of radiant barriers in consideration of the need to limit the air transfer between upper and lower air layers, these transfers leading to minimize insulation performance of the product. Finally, minimal ventilation and useful to the air layer above is provided through the ribs of the grey aluminium sheet. Simulation results show that the heat flux transmitted inside are lower than 100 W in each room, which is neglectable.

The exterior facades are covered with a clear coating (sand color). Solar absorbtivity is low and prevents overheating of exposed parts. The buffer zone is covered in wood siding facade. Its implementation is a household air gap of 40mm (cleats) that increases the thermal resistance of the wall. As for windows and bays, protection is provided by awnings, overhangs and peripheral vegetation next to the building (tall stem plants). An area of gravel is arranged around the building. This tape is a meter wide colored. To avoid heating of the air around the building, it is as much as possible shady (Picture 3)

**4.2 Window solar protection :**

Protection of windows is performed by awnings and overhangs. Facing North, an awning runs along the facade and its size was calculated using on the window height, distance to the sill and guidance. Facing West, an awning of the same type is supplemented by plants arranged at short distances. Facing east, the trees have been preserved in the landscaping fill the same role vis-à-vis the window of the loft. In everyday use, full height wooden shutters shutters are very efficient, allowing ventilation and a minimum illumination (even closed) while freeing of solar gain. Finally, the protection of stationary blinds and transoms of the mezzanine area is provided by the overhang. The size of this overhang was designed to be augmented by that of the gutter, but due to a runtime error, it was not possible to have gutters as well as expected. They were postponed after canopy, thereby increasing the shading by overhangs all openings at the bottom (living room) while significantly degrading the three bays (4 blade shutters and transom fixed) mezzanine area. Given the area involved (<1.5 m² per bay) and the volume of the living area, the thermal impact is low. Although the building is located in tropical areas, given exposure to the prevailing wind fresh in winter, these overhangs allow free valuable gains in winter (because the sun passes below) and substantial natural light in terms of visual comfort and energy saving.

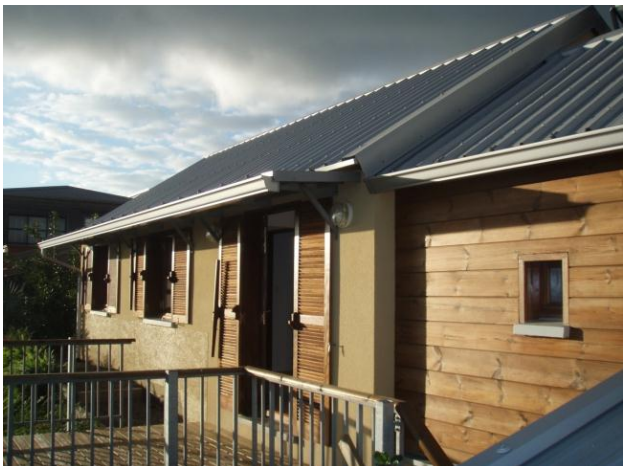

Picture 2 : South West view, night part of house

## 5. Ventilation :

Airflow simulations [6,7] have shown that the building porosity of 19% achieve the optimum dissipation of heat by ventilation. A design detail is visible on the next picture, showing the flap louvers which allow ventilation even in the closed position. The wooden shutters of Iroko wood (FSC), full height shutters, contribute day to give, even closed, a light warm atmosphere.

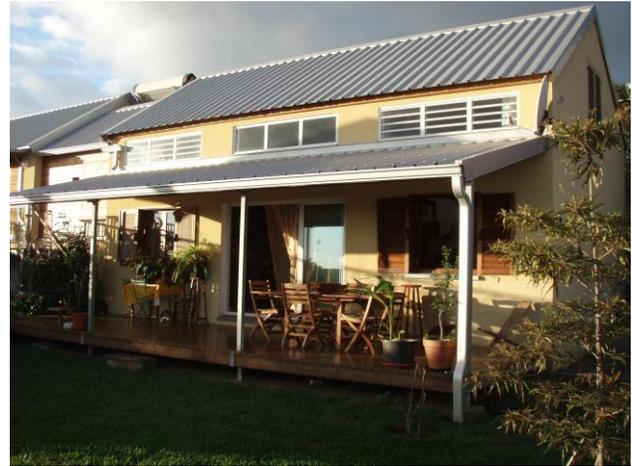

Picture 3: South West view, day part of house

Overall, for the openings, the timber has been preferred to aluminum on the project for environmental and aesthetic reasons. However, given the frequent cyclonic conditions in late austral summer, in places where certain qualities of aluminum fittings (mechanical seal) make it essential, it has been chosen.

## 6. Miscellanous :

**6.1 Renewable energy use :**

The facades and roofs are framed Northeast and exposure and is optimal, given the cloud cover currently observed during the afternoon.

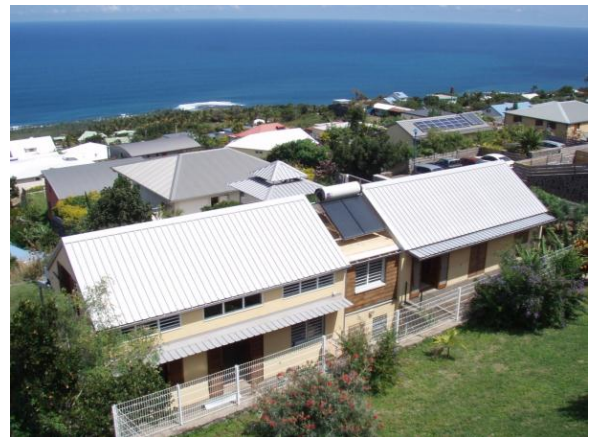

Picture 4: North view

As seen above, hot water is produced by a solar water heater of 300 liters disposed in the roof terrace. The latter was designed to allow integration into the size of the building of almost the entire system. Given the well-exposed surfaces and sloping well the absence of masks, the project has the potential to produce photovoltaic electricity, but was not affordable (10k €/kWp) in the initial budget. All electrical ducts associated with the PV system has been integrated into the masonry and is pending. The surfaces are those initially planned awnings (which would be the PV panels), the 2 roof panels inclined at 34 ° (9x4 sqm each, photo 3). Similarly, a portion of the roof of the veranda (the most southerly to minimize mask area formed by the day) could accommodate panels or a solar pool (pict. 2).

**6.2 Simulations :**
On the basis of a local weather file, the description of the building (areas, materials, ...), simulations using CODYRUN software [6-12] have firstly aimed to consolidate the assessment of comfort from the project phase, the architect being himself deeply interested in bioclimatic design. Indoor temperatures in both cases that open closed building (because CODYRUN includes a module calculating airflow through large openings), comfort indices (such as PMV) have been analyzed in detail.

As an example, the isolation of East and West walls was prescribed to be at least 2cm (polystyrene, ...). Simulations showed a performance degradation of comfort in the adjoining room where insulation. Indeed, the conductive heat flow analysis shows that these walls were ensuring the heat dissipation and their insulation tends to increase the indoor air temperatures. Only sun protection by vegetation has been maintained.

These simulations were also intended to confirm certain other technical choices. For example, at the level of roof insulation, we were asked to choose between mass insulation and radiant barrier (thin reflective insulation). The first issue was the performance of this type of insulation (in the presence of air spaces and less than a few cm).CODYRUN (because that integrating a detailed model of radiation) has validated its relevance in our case. In addition, simulations were also conducted using ECOTECT to validate aspects of lighting, as CODYRUN was not integrating at this time natural lighting aspects.

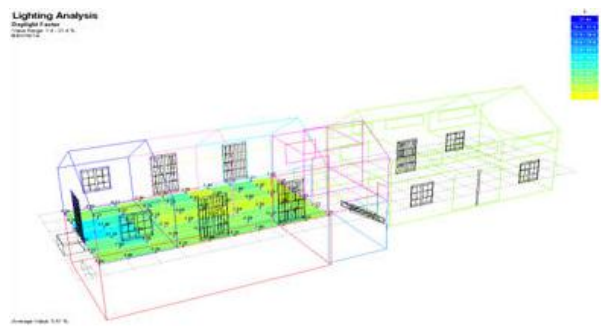

Fig 5: ECOTECT use for indoor lighting

## Conclusion
Due to passive design taken into account from the early stages, this house, subject to conditions of high sollicitations, can easily avoid artificial cooling. The used principles are very simple, such as the orientation of the building, roof insulation, use of numerous apertures, ... Moreover, its design to receive solar gains in winter makes it possible to create any throughout the year agreable indoor conditions. Finally, being open to the outdoor (windows, transoms fixed ...), the use of light is reduced and this house achieves an annual energy ratio below 20 kW/m².

## Acknowledgements
The author wishes to thank the organizers of the contest "Solar Houses, Houses Today" that have awarded a special mention to this realization [1].